\documentclass[journal=nalefd,manuscript=article]{achemso}
\usepackage{chemformula} % Formula subscripts using \ch{}
\usepackage[T1]{fontenc} % Use modern font encodings
\usepackage{amsfonts}
\usepackage{bm}
\usepackage{multirow}
\usepackage{color}
\usepackage{multirow}
\usepackage{epsfig}
\usepackage{graphicx}
\usepackage[normalem]{ulem}
\usepackage{amsmath}
\usepackage{amssymb}
\usepackage{bm}
\usepackage{dcolumn}
\usepackage{braket}
\usepackage{bbold}
\usepackage{natbib}

\usepackage{array}
\usepackage{tabulary}
\usepackage{multirow}
\newcolumntype{C}[1]{>{\centering\arraybackslash}p{#1}}
\newcolumntype{L}[1]{>{\flushleft\arraybackslash}p{#1}}
\author{Jin-Yang Li}
\affiliation{School of Physics, Northwest University, Xi'an 710127, China}
\alsoaffiliation{Institute of Modern Physics, Northwest University, Xi'an 710127, China}

\author{Yong-Kun Wang}
\affiliation{School of Physics, Northwest University, Xi'an 710127, China}

\author{Ying Zhang}
\affiliation{School of physics and astronomy, Beijing Normal University, Beijing 100875, China}

\author{Si Li}
\affiliation{School of Physics, Northwest University, Xi'an 710127, China}
\alsoaffiliation{Shaanxi Key Laboratory for Theoretical Physics Frontiers, Xi'an 710127, China}
\alsoaffiliation{Peng Huanwu Center for Fundamental Theory, Xi'an 710127, China}
\alsoaffiliation{Fundamental Discipline Research Center for Quantum Science and Technology of Shaanxi Province, Xi'an 710127, China}
\email{sili@nwu.edu.cn}

\author{Wen-Li Yang}
\affiliation{Institute of Modern Physics, Northwest University, Xi'an 710127, China}
\alsoaffiliation{Shaanxi Key Laboratory for Theoretical Physics Frontiers, Xi'an 710127, China}
\alsoaffiliation{Peng Huanwu Center for Fundamental Theory, Xi'an 710127, China}
\alsoaffiliation{Fundamental Discipline Research Center for Quantum Science and Technology of Shaanxi Province, Xi'an 710127, China}
\email{wlyang@nwu.edu.cn}

\title{Electric-Field-induced Two-Dimensional Fully Compensated Ferrimagnetism and Emergent Transport Phenomena}


\begin{document}
	
	%%%%%%%%%%%%%%%%%%%%%%%%%%%%%%%%%%%%%%%%%%%%%%%%%%%%%%%%%%%%%%%%%%%%%
	%% The "entry" environment can be used to create an entry for the
	\begin{tocentry}
		\begin{center}
			\includegraphics[width=1\textwidth]{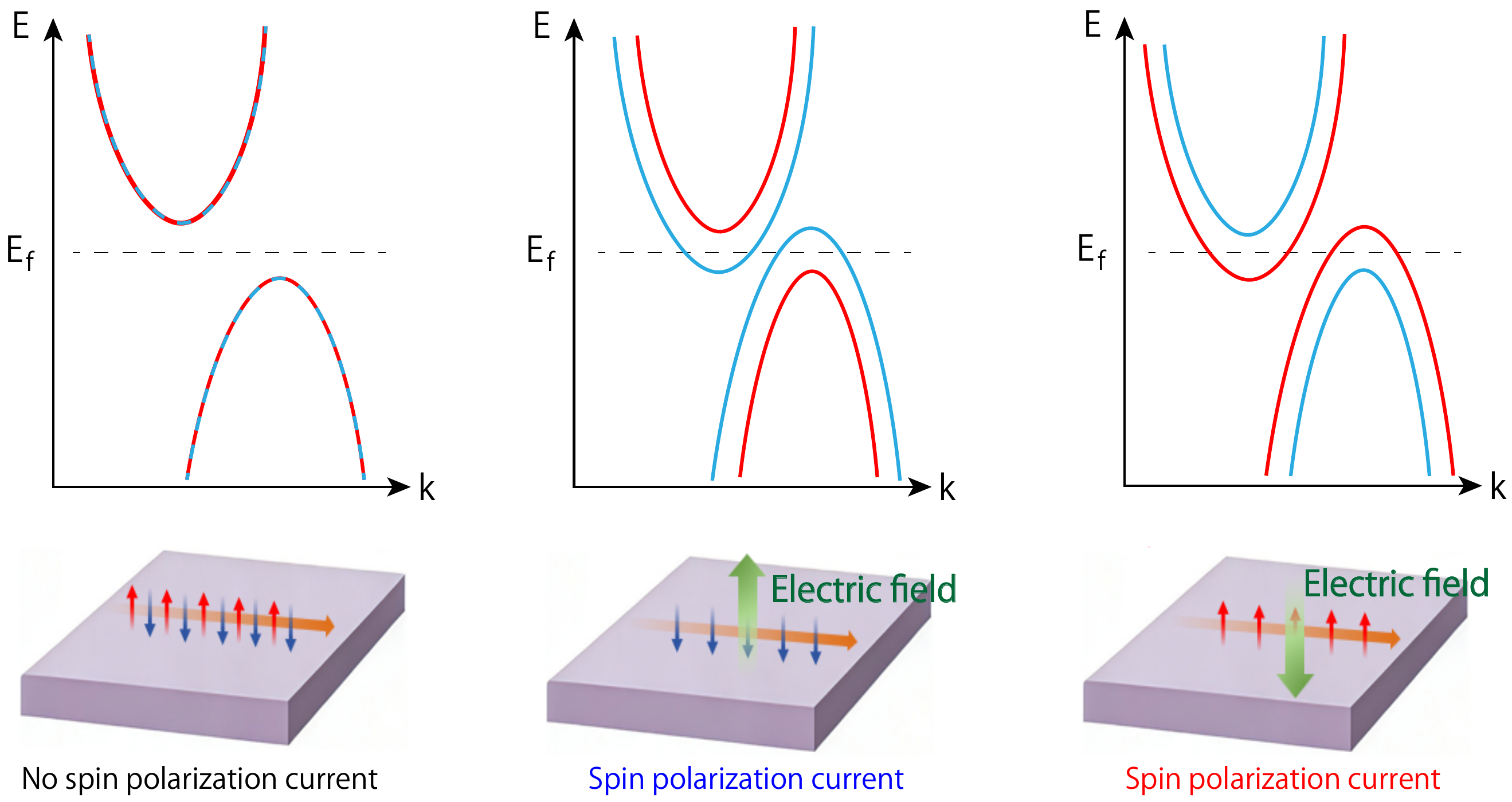}
		\end{center}
		
		%	\includegraphics[width=0.8\textwidth]{TOC_Graphic}
		%Some journals require a graphical entry for the Table of Contents.
		%This should be laid out ``print ready'' so that the sizing of the
		
		%
		%Inside the \texttt{tocentry} environment, the font used is Helvetica
		%8\,pt, as required by \emph{Journal of the American Chemical
		%The surrounding frame is 9\,cm by 3.5\,cm, which is the maximum
		%permitted for  \emph{Journal of the American Chemical Society}
		%graphical table of content entries. The box will not resize if the
		%content is too big: instead it will overflow the edge of the box.
		%
		%This box and the associated title will always be printed on a
		%separate page at the end of the document.
		%
	\end{tocentry}
	
	\newpage

\begin{abstract}
The recent discovery of altermagnetism has demonstrated that spin-split electronic band structures can emerge in magnetic systems with zero net magnetization. In contrast, fully compensated ferrimagnetic (fFIM) systems remain far less explored, despite exhibiting similar characteristics such as vanishing magnetization and spin-split bands. Here, based on first-principles calculations combined with theoretical analysis, we demonstrate that monolayer CoS and CoSe can be driven into fFIM states by an external electric field. These materials possess collinear antiferromagnetic ground states with out-of-plane Néel vectors, and their electronic bands are spin degenerate due to $\mathcal{PT}$ symmetry. When an out-of-plane electric field is applied, $\mathcal{PT}$ symmetry is broken, inducing fFIM states with pronounced spin splitting. Moreover, we show that the resulting fFIM states host fully spin-polarized currents, anomalous Hall effects, and magneto-optical Kerr and Faraday effects. Our results establish monolayer CoS and CoSe as promising platforms for electric-field-controlled fFIM states and spintronic applications.
\end{abstract}

\newpage

\begin{figure*}[htb]
	\includegraphics[width=0.95\columnwidth]{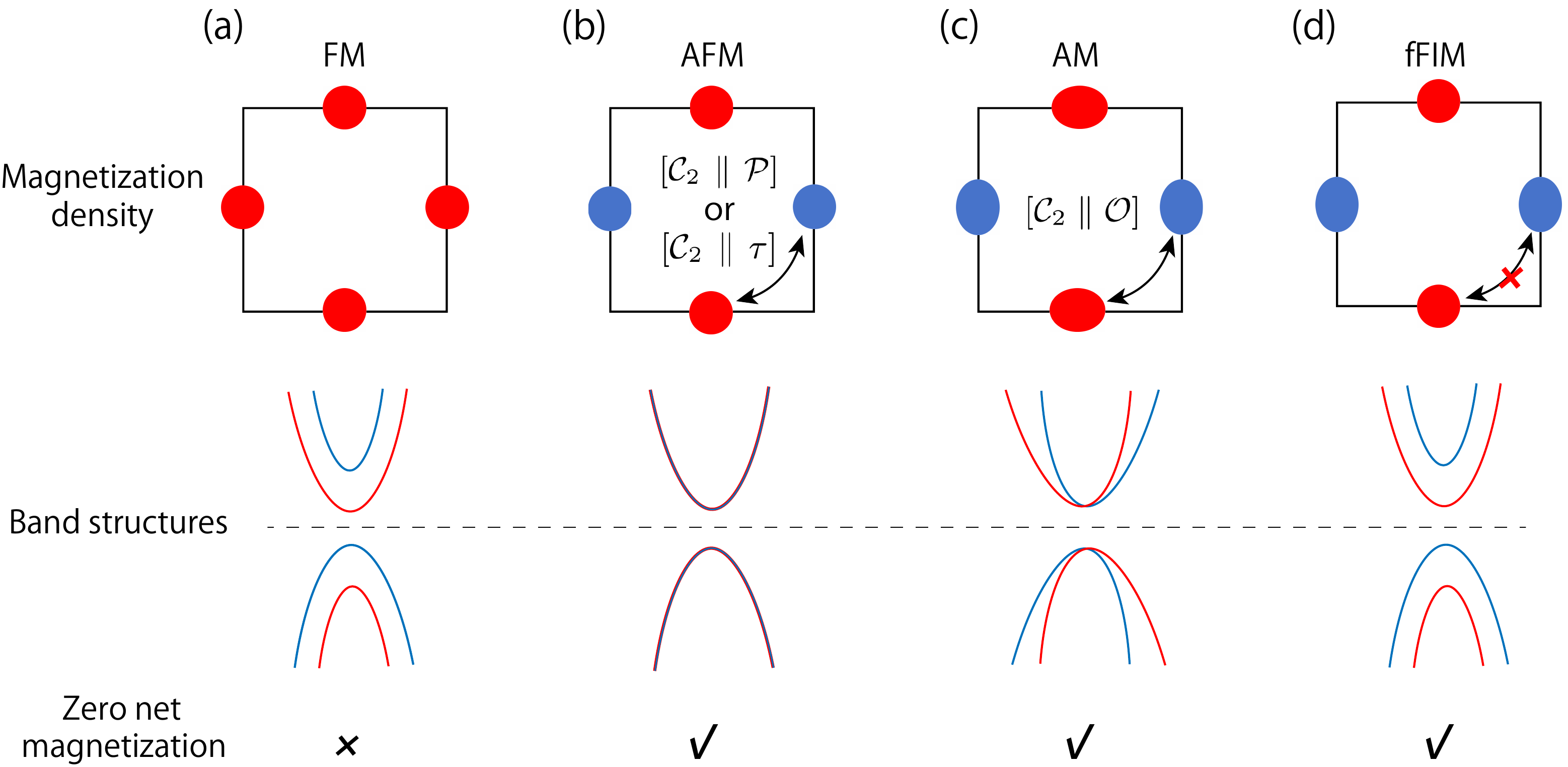}
	\caption{Four typical types of collinear magnetic order.
		(a) Ferromagnetism (FM) exhibits a finite net magnetization and spin-split bands.
		(b) Conventional antiferromagnetism (AFM) features magnetic sublattices related by $[\mathcal{C}_2 \parallel \mathcal{P}]$ or $[\mathcal{C}_2 \parallel \tau]$, resulting in zero net magnetization and spin-degenerate bands.
		(c) Altermagnetism (AM) has magnetic sublattices connected by $[\mathcal{C}_2 \parallel \mathcal{O}]$, where $\mathcal{O}$ denotes a rotation or mirror symmetry, leading to zero net magnetization with anisotropic spin splitting.
		(d) Fully compensated ferrimagnetism (fFIM) lacks symmetry relations between magnetic sublattices and exhibits zero net magnetization with isotropic spin splitting.
		Red and blue denote spin-up and spin-down sublattices or bands, respectively.}
	\label{fig1}
\end{figure*}

%\section{Introduction}
Magnetic materials play a central role in condensed matter physics, providing a fertile platform for a wide range of emergent phenomena arising from the interplay among charge, spin, and lattice degrees of freedom. Traditionally, collinear magnetism is classified into three categories: ferromagnetism, antiferromagnetism, and ferrimagnetism. Ferromagnets exhibit parallel spin alignment and spin-split electronic bands, whereas conventional antiferromagnets (AFMs) possess antiparallel spins, resulting in vanishing net magnetization and spin-degenerate bands. Ferrimagnets also possess antiparallel magnetic moments; however, their unequal magnitudes prevent complete compensation, resulting in a finite net magnetization.
Recently, a new class of magnetic materials, termed altermagnets (AMs), has been proposed~\cite{vsmejkal2022beyond,Smejkal2022a,mazin2022altermagnetism,bai2024altermagnetism,fender2025altermagnetism,song2025altermagnets}. These systems exhibit zero net magnetization while hosting anisotropic spin splitting in their electronic band structures. In addition to AFMs and AMs, another distinct class of zero-magnetization systems exists: fully compensated ferrimagnets (fFIMs)~\cite{mazin2022altermagnetism,van1995half,akai2006half,wurmehl2006valence}. Unlike AFMs and AMs, whose magnetic sublattices are connected by symmetry operations—such as space-inversion symmetry in AFMs or rotational and mirror symmetries in AMs—the magnetic sublattices in fFIMs are not connected by any symmetry operation. As a result, fFIMs exhibit ferromagnet-like spin splitting despite having zero net magnetization, as schematically illustrated in Fig.~\ref{fig1}. Thus, fFIMs represent a fundamentally distinct magnetic class beyond symmetry-protected antiferromagnetism and altermagnetism~\cite{mazin2022altermagnetism}.

The precise magnetic compensation in fFIMs is not protected by symmetry but instead originates from electron filling, which enforces an exact balance between spin-up and spin-down occupations even in the presence of localized magnetic moments~\cite{wurmehl2006valence,liu2025two}. Hence, the fFIM system exhibits nearly isotropic spin splitting across the entire Brillouin zone (BZ), giving rise to a band splitting that resembles the exchange splitting in ferromagnetic systems rather than the anisotropic spin splitting characteristic of AMs. Owing to this filling-enforced mechanism, fFIMs are robust against external perturbations such as electric fields and mechanical strain, even when such perturbations break the underlying crystal symmetries. Consequently, fFIMs exhibit physical responses closely resembling those of ferromagnets, including anomalous Hall effects, magneto-optical Kerr and Faraday effects, and fully spin-polarized currents~\cite{liu2025two}.
Previous studies of fFIMs have primarily focused on three-dimensional systems with complex crystal structures, such as double perovskite oxides, Heusler compounds, and organic materials~\cite{pickett1998spin,nie2008possible,siewierska2021magnetic,jamer2017compensated,vzic2016designing,coey2002half,hu2012half,ozdougan2015ab,stinshoff2017completely,fleischer2018magneto,kawamura2024compensated}. 
Very recently, it has been shown that fFIMs can also be realized in two-dimensional (2D) magnetic van der Waals materials through approaches such as Janus engineering, staggered potentials, and chemical substitution~\cite{liu2025two}. Despite these advances, experimentally viable routes for realizing and controlling 2D fFIMs remain limited, motivating the search for simple material platforms and efficient external control strategies.

In this work, using first-principles calculations combined with theoretical analysis, we demonstrate that monolayer CoS and CoSe host collinear Néel-type antiferromagnetic ground states with out-of-plane Néel vectors, resulting in spin-degenerate bands protected by the combined $\mathcal{PT}$ symmetry. Remarkably, an external out-of-plane electric field breaks the $\mathcal{PT}$ symmetry and drives the system into a fully compensated ferrimagnetic state with pronounced spin splitting. We further show that these electrically induced fFIM states exhibit rich emergent phenomena, including fully spin-polarized currents, anomalous Hall effects, and magneto-optical responses. Our results establish monolayer CoS and CoSe as promising platforms for electrically controllable fFIM physics and spintronic applications.

%\section{Crystal Structure, Magnetism, and Electronic Structure}
\begin{figure}[htb]
	\includegraphics[width=1\columnwidth]{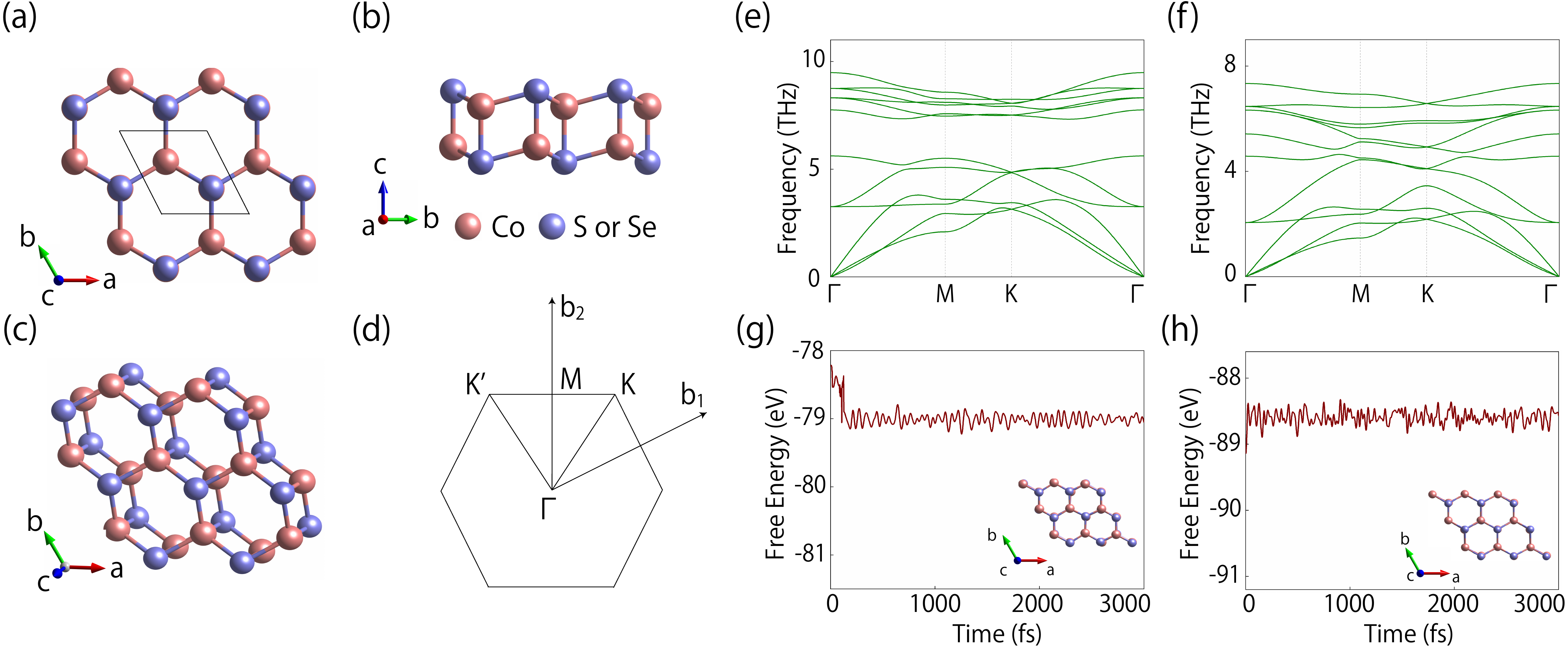}
	\caption{(a) Top, (b) side, and (c) perspective views of monolayer Co$X$ ($X$ = S, Se) crystal structure. (d) Corresponding BZ with high-symmetry points labeled. Calculated phonon spectra of monolayer (e) CoS and (f) CoSe. Ab initio molecular dynamics results of monolayer (g) CoS and (h) CoSe.}
	\label{fig2}
\end{figure}

Monolayer CoS and CoSe adopt a crystal structure similar to that of monolayer MnSe~\cite{aapro2021synthesis}. They consist of two buckled honeycomb Co$X$ ($X$ = S, Se) sublayers that are connected via Co--$X$ bonds, as illustrated in Figs.~\ref{fig2}(a)--\ref{fig2}(c). In this stacking configuration, the Co atoms in the upper sublayer alternately sit above the $X$ atoms in the lower sublayer, and vice versa. Each unit cell contains two Co atoms and two $X$ atoms. The crystal structure is centrosymmetric, belonging to the space group $P\overline{3}m1$ (No.~164) and the point group $D_{3d}$. The optimized lattice parameters of monolayer Co$X$ ($X$ = S, Se) are summarized in Table~\ref{table1}.

To examine the structural stability of these monolayers, we carried out phonon spectrum calculations and AIMD simulations. The calculated phonon spectra, shown in Figs.~\ref{fig2}(e) and \ref{fig2}(f), exhibit no imaginary phonon modes across the entire BZ, indicating dynamical stability. The AIMD results, presented in Figs.~\ref{fig2}(g) and \ref{fig2}(h), demonstrate that after 3000~fs of simulation at 300~K, the structures undergo only minor thermally induced fluctuations. Although slight lattice distortions are observed, no bond breaking or structural reconstruction occurs, confirming the excellent thermal stability of these monolayers under ambient conditions. Furthermore, these monolayer structures are inspired by experimentally synthesized layered compounds such as MnSe~\cite{aapro2021synthesis}, suggesting their strong potential for experimental realization.

%\begin{figure}[htb]
%	\includegraphics[width=0.9\columnwidth]{figure3.pdf}
%	\caption{Calculated phonon spectra of monolayer (a) CoS and (b) CoSe.
%		Ab initio molecular dynamics results of monolayer (c) CoS and (d) CoSe.}
%	\label{fig3}
%\end{figure}

\begin{table*}[htb]
	\caption{\label{table1} Calculated properties of monolayer CoS and CoSe, including the optimized lattice constant $a$ (\AA), the energy difference between Néel-type AFM and FM (zigzag-type AFM) configurations, $\Delta E_{\mathrm{NAFM-FM}}$ ($\Delta E_{\mathrm{NAFM-ZAFM}}$) (eV per primitive cell), the exchange parameter $J_{1}$ and $J_{2}$ (meV), the magnetocrystalline anisotropy energy (MAE) with SOC included--defined as the energy difference between the [001] and [100] directions, $\Delta E_{\mathrm{001-100}}$ ($\mu$eV per primitive cell), and the Néel temperature $T_{N}$ (K).}
		\begin{tabular}{cccccccc}
			\hline\hline
			Systems & $a$   & $\Delta E_{\mathrm{NAFM-FM}}$  & $\Delta E_{\mathrm{NAFM-ZAFM}}$ & $J_{1}$ & $J_{2}$ & $\Delta$$E_{001-100}$  & $T_{N}$ \\
			\hline
			CoS & 3.85 &  $-0.671$ & $-0.168$ & $-112.6$  & $35.3$  & $-240.5$   & $400$ \\
			CoSe &4.03 &  $-0.624$ & $-0.135$ & $-104.2$  & $35.1$  & $-132.9$   & $390$ \\			
		\hline\hline
		\end{tabular}
\end{table*}

As a $3d$ transition-metal element, cobalt naturally gives rise to magnetic ordering. To identify the magnetic ground state of monolayer CoS and CoSe, we systematically compared the total energies of three representative magnetic configurations, namely the ferromagnetic (FM), Néel-type antiferromagnetic (NAFM), and zigzag-type antiferromagnetic (ZAFM) states (see the Supporting Information). Our calculations reveal that both compounds energetically favor the Néel-type AFM configuration [see Fig.S1 (b)]. The corresponding energy differences between these magnetic states are summarized in Table~\ref{table1}. In this Néel-type AFM ground state, the magnetic moments are mainly localized on the Co sites, with each Co atom carrying a moment of approximately $2.3~\mu_B$.
We further estimated the Néel temperature $T_N$ of the Néel-type AFM ground state using Monte Carlo simulations based on an effective spin Hamiltonian~\cite{evans2014atomistic},
\begin{equation}\label{Heisenberg}
	H=-\sum_{i \neq j} J_{i j} \bm{S}_{i} \cdot \bm{S}_{j}-\frac{k_{N}}{2} \sum_{i}\left(S_{i}^{z}\right)^{2},
\end{equation}
where $i$ and $j$ label the Co atomic sites, $J_{ij}$ denotes the exchange interaction strength, and $k_N$ represents the magnetic anisotropy constant. All model parameters were extracted from first-principles calculations. The obtained nearest-neighbor exchange interaction $J_1$, next-nearest-neighbor interaction $J_2$, and magnetic anisotropy strength $k_N$ are listed in Table~\ref{table1}. The Néel temperature is determined from the temperature dependence of the mean sublattice magnetization, as shown in Fig.S1 (d) and~Fig.S1 (e). Based on the MC simulations, the estimated $T_N$ values are approximately 400~K for CoS and 390~K for CoSe. These relatively high Néel temperatures indicate that monolayer CoS and CoSe are promising candidate for practical spintronic applications.

\begin{figure*}[htb]
	\includegraphics[width=0.95\columnwidth]{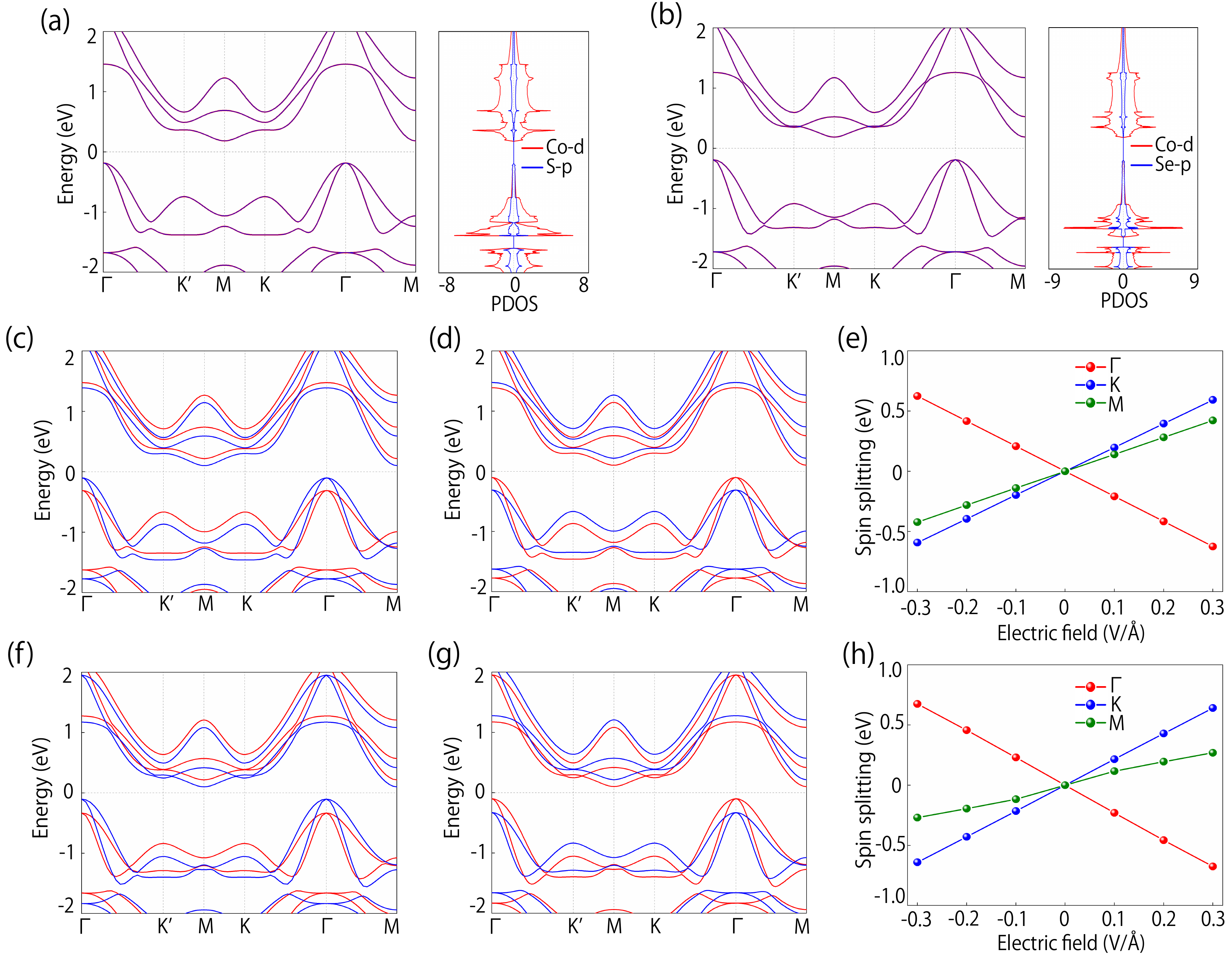}
	\caption{Band structures and projected density of states of monolayer (a) CoS and (b)CoSe in the absence of SOC. Band structures of monolayer CoS under out-of-plane electric fields of (c) $E = 0.1$ V/Å and (d) $E = -0.1$ V/Å, and (e) the corresponding spin splitting of the top valence band at $\Gamma$, K, and M.
		Band structures of monolayer CoSe under out-of-plane electric fields of (f) $E = 0.1$ V/Å and (g) $E = -0.1$ V/Å, and (h) the corresponding spin splitting of the top valence band at $\Gamma$, K, and M.}
	\label{fig3}
\end{figure*}

We calculated the electronic band structures of monolayer CoS and CoSe in their ground state, namely the Néel-type AFM configuration. The calculated band structures and projected density of states (PDOS) without spin--orbit coupling (SOC) are shown in Figs.~\ref{fig3}(a) and~\ref{fig3}(b). As can be seen, both materials exhibit spin-degenerate energy bands and are identified as indirect band-gap semiconductors. The spin degeneracy of the energy bands arises from the preserved combined $\mathcal{PT}$ symmetry in the Néel-type AFM ground state. The conduction band minimum (CBM) is located at the M point, while the valence band maximum (VBM) appears at the $\Gamma$ point. The corresponding global band gaps are 0.367 eV for CoS and 0.376 eV for CoSe. The PDOS indicates that the low-energy states near the valence band are mainly derived from Co $d$ orbitals. In addition, the band structures exhibit well-defined valleys at the $K$ and $K'$ points in the valence band.

%\section{RESULTS}
%\section{Electric-Field-Induced Fully Compensated Ferrimagnetism}

In the following, we investigate electric-field-induced fully compensated ferrimagnetism in monolayer CoS and CoSe. As discussed above, the ground state of monolayer CoS and CoSe adopts a Néel-type AFM configuration that preserves the combined $\mathcal{PT}$ symmetry. To achieve fully compensated ferrimagnetism, this $\mathcal{PT}$ symmetry must be broken while maintaining a vanishing total magnetic moment. An applied out-of-plane electric field breaks the inversion symmetry $\mathcal{P}$ by inducing a layer-dependent electrostatic potential, thereby breaking the $\mathcal{PT}$ symmetry and giving rise to fully compensated ferrimagnetism. The band structures of monolayer CoS and CoSe under an out-of-plane electric field of 0.1 V/Å are shown in Figs.~\ref{fig3}(c) and~\ref{fig3}(f), respectively. As can be seen from Figs.~\ref{fig3}(c) and~\ref{fig3}(f), a pronounced ferromagnet-like spin splitting emerges in the band structure, originating from the breaking of the $\mathcal{PT}$ symmetry induced by the applied electric field. We find that upon application of the electric field, the magnetic moments of the two Co atoms remain equal in magnitude and opposite in direction (each still being $ 2.3~\mu_{\mathrm{B}}$), resulting in a vanishing total magnetic moment. Consequently, the electric field drives monolayer CoS and CoSe into a fully compensated ferrimagnetic state. Interestingly, reversing the direction of the applied electric field also reverses the spin splitting of the energy bands, as illustrated in Figs.~\ref{fig3}(d) and~\ref{fig3}(g). To investigate the relationship between the spin-splitting magnitude and the applied electric field, we plot the energy differences between the two highest valence bands at the $\Gamma$, $K$, and $M$ points (defined as $E_\uparrow - E_\downarrow$) in Figs.~\ref{fig3}(e) and~\ref{fig3}(h). The results show that the spin splitting increases nearly linearly with increasing electric field strength.

 Experimentally, the magnetic configuration of the fFIM phase can be probed by neutron scattering or x-ray magnetic linear dichroism~\cite{shull1949detection,luning2003determination}. The ferromagnet-like spin splitting in the band structure of the fFIM phase can be detected using spin-resolved angle-resolved photoemission spectroscopy~\cite{zhu2024observation}.

%\section{Fully Spin-Polarized Currents}
\begin{figure*}[htb]
	\includegraphics[width=1\columnwidth]{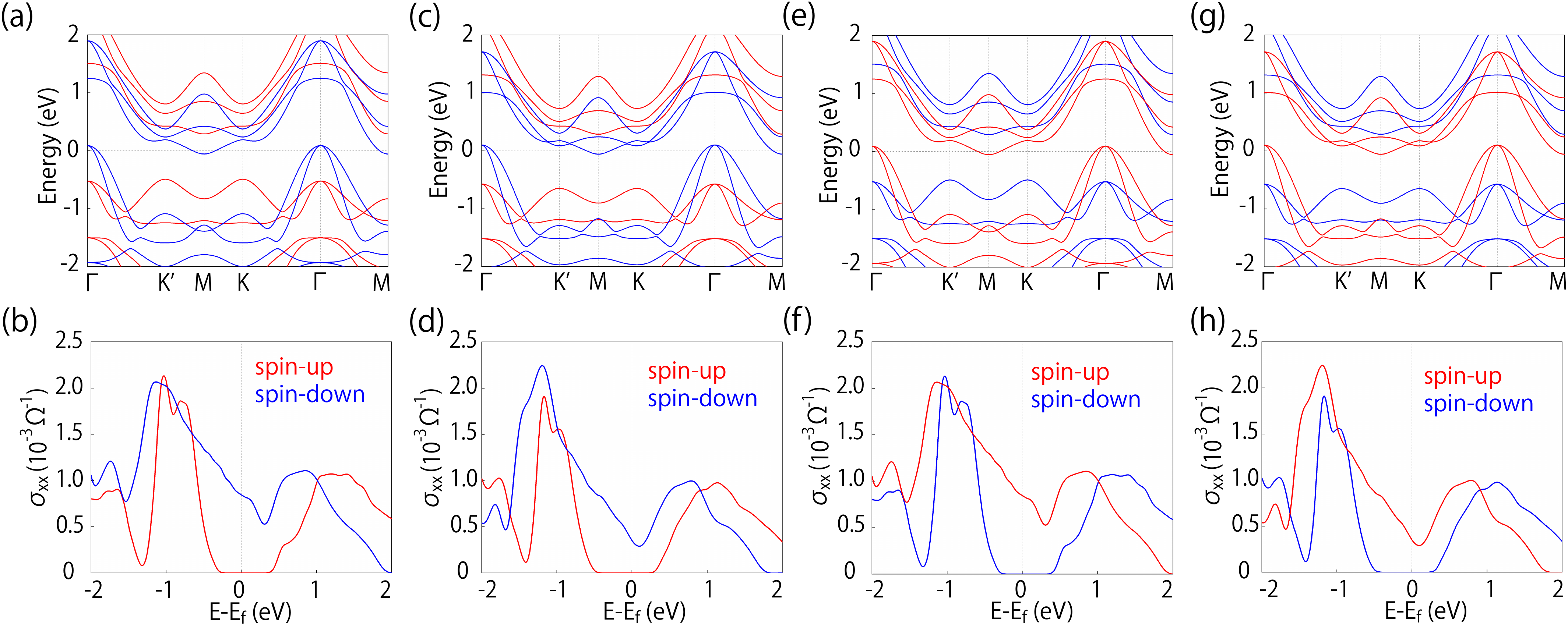}
	\caption{Band structures of monolayer (a) CoS and (c) CoSe under an out-of-plane electric field of $E = 0.3$ V/Å. Spin-resolved longitudinal conductivity $\sigma_{xx}$ of monolayer (b) CoS and (d) CoSe. Band structures of monolayer (e) CoS and (g) CoSe under an out-of-plane electric field of $E = -0.3$ V/Å. Corresponding spin-resolved longitudinal conductivity $\sigma_{xx}$ for monolayer (f) CoS and (h) CoSe.
	}
	\label{fig4}
\end{figure*}

Next, we investigate the spin-polarized currents in these fully compensated ferrimagnetic systems under an external electric field. We find that when the field exceeds a critical threshold, the induced spin splitting becomes large enough to drive the system from a semiconducting state into a spin-polarized metallic phase, where only a single spin channel crosses the Fermi level. For instance, at an applied electric field of 0.3 V/Å, the calculated band structures of monolayer CoS and CoSe are shown in Figs.~\ref{fig4}(a) and~\ref{fig4}(c), respectively. As can be seen, the global band gap closes and both systems become metallic. Notably, only spin-down bands appear near the Fermi level, indicating the emergence of a spin-polarized metallic state. Such spin-polarized metals are capable of generating fully spin-polarized charge currents.
In the absence of SOC, spin transport is characterized by two independent conductivity components, $\sigma_{xx}^\uparrow$ and $\sigma_{xx}^\downarrow$. Owing to the threefold rotational symmetry $C_{3z}$, the in-plane conductivities satisfy $\sigma_{yy}^\uparrow=\sigma_{xx}^\uparrow$ and $\sigma_{yy}^\downarrow=\sigma_{xx}^\downarrow$. The calculated spin-resolved charge conductivities $\sigma_{xx}^\uparrow$ and $\sigma_{xx}^\downarrow$ of monolayer CoS and CoSe are presented in Figs.~\ref{fig4}(b) and~\ref{fig4}(d), respectively. It is evident that within a finite energy window around the Fermi level, the spin-up conductivity vanishes, whereas the spin-down conductivity remains finite, thereby enabling fully spin-polarized current transport. Moreover, reversing the direction of the electric field can switch the spin polarization of the electronic states near the Fermi level [see Figs.~\ref{fig4}(e)--\ref{fig4}(h)], thereby enabling efficient electrical control of spin-polarized transport.

\begin{figure}[htb]
	\includegraphics[width=1\columnwidth]{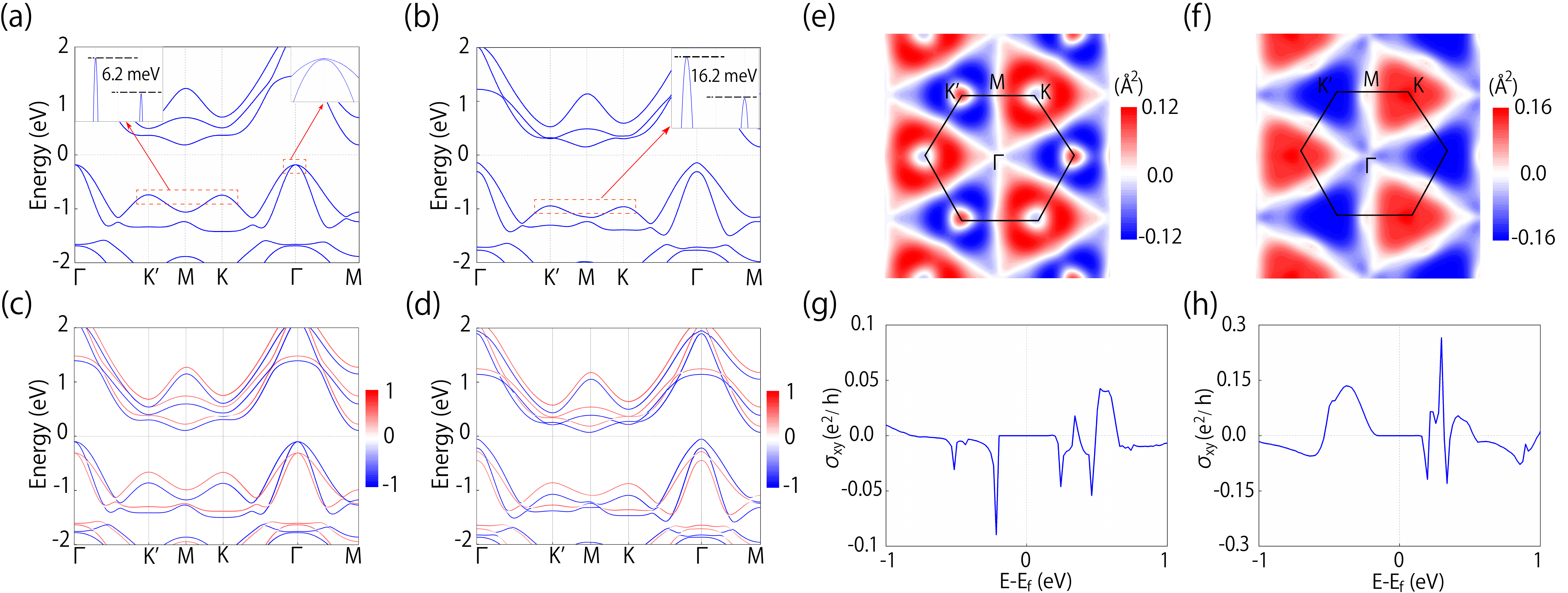}
	\caption{Electronic band structures with SOC of monolayer (a) CoS and (b) CoSe. Insets show a small band gap at the valence band maximum at the $\Gamma$ point and the lifting of valley degeneracy at the $K$ and $K'$ points.
		Band structures with spin projection $s_z$ of monolayer (c) CoS and (d) CoSe under an out-of-plane electric field of $E = 0.1$ V/Å. Berry curvature distribution summed over all valence bands of (e) CoS and (f) CoSe under an out-of-plane electric field of $E = 0.1$ V/Å. Anomalous Hall conductivity of monolayer (g) CoS and (h) CoSe under the same electric field.}
	\label{fig5}
\end{figure}

%\section{Electronic Band Structure with SOC and Anomalous Hall Effect}
Another notable feature of fully compensated ferrimagnetic systems is their ability to host a nonzero Berry curvature and exhibit a finite anomalous Hall conductivity, despite having zero net magnetization.
We first examine the orientation of the Néel vector and the electronic band structures of monolayer CoS and CoSe, including SOC. The magnetocrystalline anisotropy energy (MAE) is computed by aligning the magnetization along both in-plane and out-of-plane directions and comparing the corresponding total energies. Our results indicate that the out-of-plane direction ([001]) serves as the magnetic easy axis, while the in-plane directions are nearly isotropic in energy. Accordingly, the MAE is defined as the energy difference between the [001] and [100] directions, with the calculated values summarized in Table~\ref{table1}.
In the absence of an external electric field, the electronic band structures of monolayer CoS and CoSe in the $z$-direction ground state, including SOC, are shown in Figs.~\ref{fig5}(a) and~\ref{fig5}(b), respectively. It can be seen that SOC has only a minor effect on the overall band structures, and the bands remain spin degenerate. Notably, a small band gap appears at the valence-band maximum at the $\Gamma$ point. Additionally, the valley degeneracy of the highest valence band at the $K$ and $K'$ points is lifted, as highlighted in the enlarged views.
Upon applying an external electric field, the spin-projected band structures with SOC are shown in Figs.~\ref{fig5}(c) and~\ref{fig5}(d). The results clearly reveal pronounced spin splitting of the bands induced by the electric field.

We then investigate the Berry curvature and the anomalous Hall effect. In 2D systems, the Berry curvature has only a single nonzero component along the out-of-plane ($z$) direction. The Berry curvature is defined as
\begin{equation}\label{BC}
		\Omega_{z}\left(\bm k\right)=-2 \operatorname{Im} \sum_{n\neq n^{\prime}}f_{n\bm k} \frac{\left\langle n \bm{k}\left|v_{x}\right| n' \bm{k}\right\rangle\left\langle n' \bm{k}\left|v_{y}\right| n \bm{k}\right\rangle}{(\omega_{n^{\prime}}-\omega_{n})^{2}},
\end{equation}
where $v_{x/y}$ are the velocity operators, and $E_n=\hbar\omega_{n}$ is the energy of the state $|n\bm k\rangle$. In the absence of the out-of-plane electric field, the system preserves $\mathcal{PT}$ symmetry, which enforces the Berry curvature to vanish at every $\bm k$ point in the BZ. Consequently, the Berry curvature is identically zero and no anomalous Hall response is expected. When the electric field is applied, the $\mathcal{PT}$ symmetry is broken, driving the system into the fFIM phase and allowing a finite Berry curvature to emerge. The total Berry curvature summed over all occupied valence bands is shown in Figs.~\ref{fig5}(e) and~\ref{fig5}(f) for monolayer CoS and CoSe, respectively. The distribution exhibits pronounced peaks at the $K$ and $K'$ valleys with opposite signs, highlighting the valley-contrasting nature of the Berry curvature. This nonzero Berry curvature gives rise to anomalous Hall transport, where an in-plane electric field induces a transverse carrier velocity even in the absence of an external magnetic field. We then calculate the intrinsic anomalous Hall conductivity, a fundamental quantity determined solely by the geometric properties of the electronic band structure. This quantity can be accurately evaluated through first-principles calculations~\cite{jungwirth2002anomalous,yao2004first} and is expressed as
\begin{equation}
	\sigma_{x y}^i=-\frac{e^{2}}{\hbar} \int_\text{BZ} \frac{d^2 k}{(2\pi)^2} \Omega_z\left(\bm k\right),
\end{equation}
where $\Omega_z\left(\bm k\right)$ is the $z$-component of the total Berry curvature of the occupied states at $\bm k$, as defined in Eq.~(\ref{BC}). The intrinsic anomalous Hall conductivity obtained from our calculations is shown in Figs.~\ref{fig5}(g) and ~\ref{fig5}(h), with peak values comparable to those of typical transition-metal ferromagnets.

%\section{Magneto-Optical Effects}
\begin{figure*}[htb]
	\includegraphics[width=1\columnwidth]{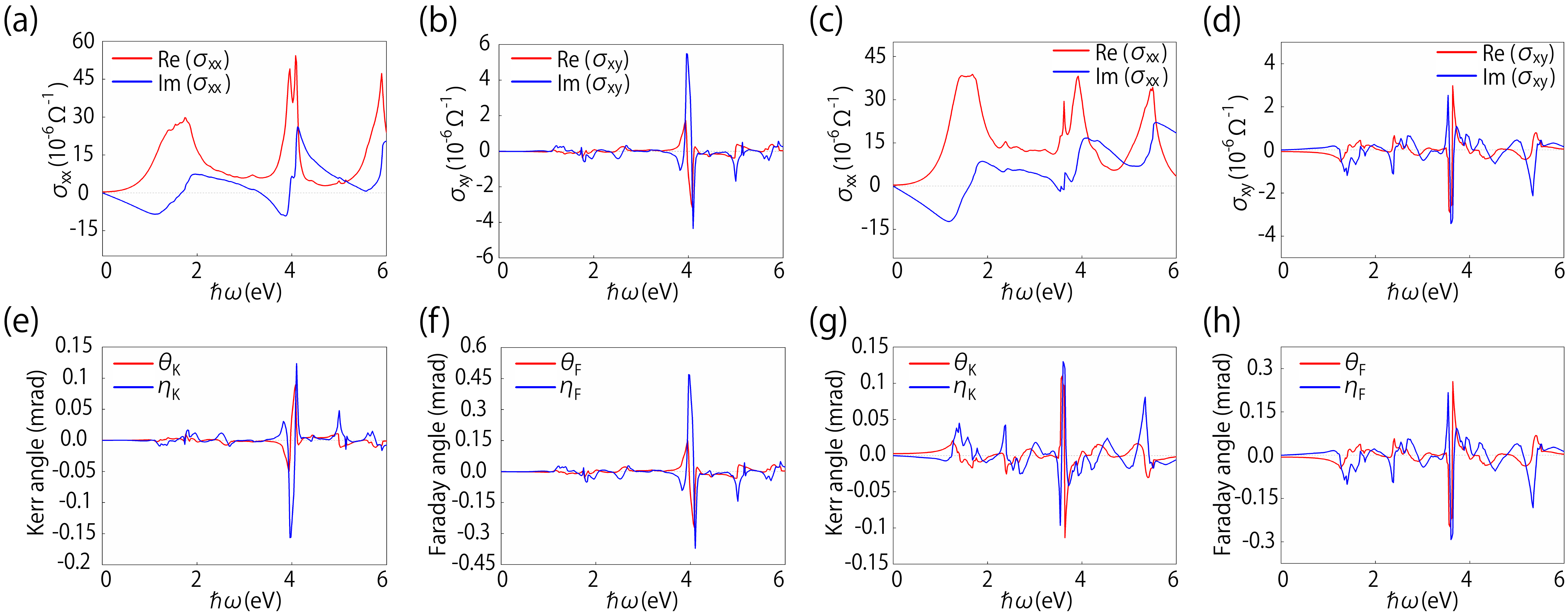}
	\caption{(a) Calculated diagonal component ($\sigma_{xx}$) and (b) off-diagonal component ($\sigma_{xy}$) of the optical conductivity tensor for monolayer CoS under an out-of-plane electric field of 0.1 V/Å.
		(c) Calculated diagonal component ($\sigma_{xx}$) and (d) off-diagonal component ($\sigma_{xy}$) of the optical conductivity tensor for monolayer CoSe under the same electric field.
		(e) Kerr angle and (f) Faraday angle for monolayer CoS, and (g) Kerr angle and (h) Faraday angle for monolayer CoSe under an out-of-plane electric field of 0.1 V/Å.}
	\label{fig6}
\end{figure*}

In addition to exhibiting fully spin-polarized currents and anomalous Hall effects, the fFIM phase also displays pronounced magneto-optical responses, including Kerr and Faraday effects. The complex Kerr and Faraday angles in two-dimensional systems~\cite{kim2007determination,valdes2012terahertz} can be expressed as
\begin{align}  
\theta_{K}+i\eta_{K}&=\frac{2(Z_{0}d\sigma_{xy})}{1-(n_{s}+Z_{0}d\sigma_{xx})^{2}},\nonumber\\
\theta_{F}+i\eta_{F}&=\left(\frac{\sigma_{xy}}{\sigma_{xx}}\right)\left[1+\frac{n_{s}+1}{Z_{0}d\sigma_{xx}}\right]^{-1},
\end{align}
where \(\theta\) and \(\eta\) denote the rotation angle and ellipticity, respectively.
Here, $\sigma_{xx}$ and $\sigma_{xy}$ are the diagonal and off-diagonal components of the optical conductivity tensor, characterizing the material’s response to an applied electromagnetic field. $Z_{0}$ is the free-space impedance ($\sim 377~\Omega$), $d$ is the sample thickness, and $n_{s}$ is the refractive index of the substrate. In this work, $n_{s}$ is taken to be 1.5, corresponding to SiO$_2$.
The optical conductivity tensor $\boldsymbol{\sigma}$ is calculated using the Kubo–Greenwood formula~\cite{yates2007spectral}
\begin{eqnarray}\label{eq:OPC}
	\sigma_{\mu\nu}&=& \frac{ie^2\hbar}{N_k V}\sum_{\textbf{k}}\sum_{n, m}\frac{f_{m\textbf{k}}-f_{n\textbf{k}}}{E_{m\textbf{k}}-E_{n\textbf{k}}} \frac{\langle\psi_{n\textbf{k}}|\hat{\upsilon}_{\mu}|\psi_{m\textbf{k}}\rangle\langle\psi_{m\textbf{k}}|\hat{\upsilon}_{\nu}|\psi_{n\textbf{k}}\rangle}{E_{m\textbf{k}}-E_{n\textbf{k}}-(\hbar\omega+i\eta)},
\end{eqnarray}
where $V$ is the volume of the unit cell, $N_k$ is the total number of $\bm{k}$ points in the BZ, $\omega$ is the photon frequency, and $\eta$ is the energy-smearing parameter, $f_{n\bm{k}}$ denotes the Fermi–Dirac distribution function. $\hat{\upsilon}_{\mu(\nu)}$ is the velocity operator, where the subscripts $\mu,\nu \in \{x,y,z\}$ label the Cartesian components. $\psi_{n\bm{k}}$ and $E_{n\bm{k}}$ are the Wannier functions and the interpolated energy at band index $n$ and momentum $\bm{k}$, respectively.
Using this framework, we evaluate the Kerr and Faraday responses of monolayer CoS and CoSe under an out-of-plane electric field of 0.1 V/Å, as shown in Fig.~\ref{fig6}. From Fig.~\ref{fig6}, one can observe that the maximum Kerr rotation angle is comparable in magnitude to that of monolayer CrI$_3$~\cite{huang2017layer}, confirming a significant magneto-optical effect in these systems.

%\section{CONCLUSION}
In conclusion, based on first-principles calculations and theoretical analysis, we have demonstrated that monolayer CoS and CoSe can host electrically tunable fully compensated ferrimagnetic states. In their ground states, both materials exhibit collinear Néel-type antiferromagnetic ordering with spin-degenerate electronic bands protected by combined $\mathcal{PT}$ symmetry. The application of an out-of-plane electric field breaks this symmetry and induces fFIM states characterized by pronounced spin splitting while maintaining zero net magnetization. Moreover, these fFIM states support a series of intriguing physical effects, including fully spin-polarized currents, anomalous Hall responses, and magneto-optical Kerr and Faraday effects. Our findings not only enrich the understanding of fully compensated magnetism in two-dimensional systems but also highlight monolayer CoS and CoSe as attractive candidates for future spintronic devices controlled by electric fields.

\begin{suppinfo}	
	The Supporting Information contains: first-principles methods, magnetic configurations and
	band structure results with different $U$ values.
\end{suppinfo}	

\bigskip
	\begin{acknowledgement}
The authors thank A. D. Fan for helpful discussions. This work was supported by the National Natural Science Foundation of China (Grants No. 12204378, No. 12434006, and No. 12247103), the Key Program of the Natural Science Basic Research Plan of Shaanxi Province (Grant No. 2025JC-QYCX-007), and the Major Basic Research Program of Natural Science of Shaanxi Province (Grant No. 2021JCW-19).
	\end{acknowledgement}

\bibliography{CoX_ref}

\end{document}